\documentclass
[byrevtex,showpacs,balancelastpage,10pt,twocolumn,a4paper,amsfonts,amssymb,preprint,prb,floats,showkeys]{revtex4}%
\usepackage{amsfonts}
\usepackage{amsmath}
\usepackage{amssymb}
\usepackage[dvips]{graphicx}%
\providecommand{\U}[1]{\protect\rule{.1in}{.1in}}
\providecommand{\U}[1]{\protect\rule{.1in}{.1in}}
\providecommand{\U}[1]{\protect\rule{.1in}{.1in}}
\newtheorem{theorem}{Theorem}
\newtheorem{acknowledgement}[theorem]{Acknowledgement}

\ifx\pdfoutput\relax\let\pdfoutput=\undefined\fi
\newcount\msipdfoutput
\ifx\pdfoutput\undefined\else
\ifcase\pdfoutput\else
\ifx\paperwidth\undefined\else
\ifdim\paperheight=0pt\relax\else\pdfpageheight\paperheight\fi
\ifdim\paperwidth=0pt\relax\else\pdfpagewidth\paperwidth\fi
\fi\fi\fi
\begin{document}
\title{Intrinsic electron-glass effects in strongly-localized thallium-oxide films}
\author{Z. Ovadyahu}
\affiliation{Racah Institute of Physics, The Hebrew University, Jerusalem, 91904, Israel}

\begin{abstract}
Transport measurements made on films of thallium-oxide (n-type semiconductor)
are presented and discussed. The focus in this work is on the
strongly-localized regime where charge transport is by variable-range-hopping.
It is demonstrated that, at liquid-helium temperatures, these films exhibit
all the characteristic features of intrinsic electron-glasses. These include a
slow (logarithmic in time) conductance-relaxation that may be induced by any
of the following protocols: Quench-cooling from high temperatures, sudden
change of gate-voltage, exposure to infrared radiation, and stressing the
system with a non-Ohmic field. The microstructure of the films are
characterized by electron microscopy and their carrier-concentration are
measured by Hall effect. Field-effect experiments reveal a memory-dip that has
a width compatible with the carrier-concentration of the system as compared
with previously studied electron-glasses. It is observed that the common
ingredient in all the systems that exhibit electron-glass effects is high
carrier-concentration suggesting that their localized sites may be
multi-occupied even when deep into the insulating regime. That lightly-doped
semiconductors do not show intrinsic electron-glass effects is consistent with
this empirical observation. The connection between the memory-dip and the
Coulomb-gap is discussed in light of these findings.

\end{abstract}

\maketitle

\section{Introduction}

The interplay between disorder and interactions in Fermi systems has been a
challenging field of research and a source of fascination for more than five
decades. A prominent representative of this group, the Anderson transition,
has been extensively studied both theoretically and experimentally, and at
least qualitatively, its equilibrium properties are well understood. On the
other hand, the out-of-equilibrium properties of the insulating phase remains
ill-understood and controversial. This applies in particular to the
`electron-glass' (EG) scenario discussed in several papers \cite{1,2,3,4,5}.
The EG phase is expected to arise from the competition between disorder and
interactions. It is a generic scenario that applies to all degenerate Fermi
systems with localized states interacting via a Coulomb potential. It ought to
be observed in any strongly-localized system exhibiting hopping transport.
Experimental evidence for glassy effects, however, has been somewhat scarce.
Intrinsic EG effects were found in several systems but remarkably none was
reported in lightly-doped semiconductors. By `intrinsic EG' we mean that the
effects appear systematically for a given substance independently of the way
the sample was prepared to achieve the same relevant parameters (resistance at
the measuring temperature, dimensionality, and carrier-concentration).

It has been conjectured \cite{6} that the reason for the absence of intrinsic
EG effects in lightly-doped semiconductors is their inherently short
relaxation-time, which makes it hard to resolve these effects by transport
measurements. The short relaxation time of semiconductors was alleged to be
due to their low carrier-concentrations n \cite{6}. This has been supported by
a recent experiment on phosphorous-doped silicon where ultra-fast relaxation
was found \cite{7}. Conductance relaxations that persist for many seconds, and
memory effects characteristic of intrinsic electron-glass, seem \cite{6,8} to
be peculiar to systems with n$\curlyeqsucc$5$\cdot$10$^{\text{19}}%
$cm$^{\text{-3}}$.

Intrinsic electron-glass effects were observed in films of crystalline
indium-oxide \cite{9} (In$_{\text{2}}$O$_{\text{3-x}}$), several versions of
amorphous indium-oxide \cite{9} (In$_{\text{x}}$O), and, in ultra-thin films
of beryllium \cite{10}. In all these materials the carrier-concentration was
measured by Hall effect and verified to be in the high-n limit.

In this work we report on the low temperature transport properties of
thallium-oxide, which is n-type semiconductor with high carrier-concentration.
The microstructure of thallium-oxide films is quite different than those of
In$_{\text{2}}$O$_{\text{3-x}},$ In$_{\text{x}}$O, and beryllium. However,
strongly-localized films of this material exhibit glassy effects that are
essentially identical to those of In$_{\text{2}}$O$_{\text{3-x}},$
In$_{\text{x}}$O, and beryllium. These include logarithmic relaxation of the
out-of-equilibrium conductance and a memory-dip (MD) that has all the earmarks
of intrinsic electron-glass. The width of the memory-dip exhibited by
thallium-oxide is observed to be close to that of the electron-rich version of
In$_{\text{x}}$O that has similar carrier-concentration. This appears to be in
line with a general trend; the MD width of all materials tested so far shows a
systematic correlation with carrier-concentration. The possible connection
between the memory-dip and the Coulomb-gap, and the origin of similar behavior
in granular systems are discussed in light of these findings.

\section{Experimental}

\subsection{Sample preparation and characterization}

The Tl$_{\text{2}}$O$_{\text{3-x}}$ films used here were e-gun evaporated on
room-temperature microscope-slides or, for the field effect measurements, unto
the 0.5$\mu$m SiO$_{\text{2}}$ layer thermally grown on
%TCIMACRO{\TEXTsymbol{<}}%
%BeginExpansion
$<$%
%EndExpansion
100%
%TCIMACRO{\TEXTsymbol{>} }%
%BeginExpansion
$>$
%EndExpansion
Si-wafers. The Si wafers were boron-doped (P-type) with bulk resistivity
$\rho\simeq$ 2$\cdot$10$^{\text{-3}}\Omega$-cm deeply into the degenerate
regime. They were used in this study as the gate electrode in the field-effect
experiments. The substrate was rinsed by ethanol, dried, and was ion-bombarded
by O$_{\text{2}}$ in the vacuum chamber prior to material evaporation. The
source material was 99\% pure Tl$_{\text{2}}$O$_{\text{3}}$ brown powder
(American Elements) pressed into the water-cooled copper crucible of the e-gum
ficture. Deposition was carried out at (0.5-2)$\cdot$10$^{\text{-4}}$ Torr
oxygen ambience maintained by leaking 99.9\% pure O$_{2}$ through a needle
valve into the vacuum chamber (base pressure $\simeq$10$^{\text{-6}}$ Torr).
Rates of deposition used for the samples reported here were in the range
0.2-4~\AA /s monitored in-situ using a quartz thickness-monitor calibrated by
a Tolansky interferometer.

The Tl$_{\text{2}}$O$_{\text{3}}$ powder temperature during deposition was
600$\pm$30K, a much lower evaporation temperature than the $\simeq$1500K used
for In$_{\text{2}}$O$_{\text{3-x}}$ to obtain similar evaporation rates
\cite{11}. Another aspect where Tl$_{\text{2}}$O$_{\text{3-x}}$ films differ
from In$_{\text{2}}$O$_{\text{3-x}}$ films is in the way they respond to
different post-preparation treatment. In particular, exposing In$_{\text{2}}%
$O$_{\text{3-x}}$ to UV radiation breaks oxygen bonds, which upon vacuuming
are removed from the sample. This in turn increases the carrier-concentration
and decreases the sample resistance \cite{9}. The UV treatment is a reversible
process by which a given sample resistance may be changed. It works well in
In$_{\text{2}}$O$_{\text{3-x}}$ films as well as in ZnO crystals \cite{12}
where the resistance may be restored by letting oxygen diffuse back into the
sample. Films of Tl$_{\text{2}}$O$_{\text{3-x}}$ on the other hand, show no
change of resistance when exposed to UV radiation, and are much less
susceptible to oxidizing agents than In$_{\text{2}}$O$_{\text{3-x}}$. Lacking
this flexibility of fine-tuning the disorder, thallium-oxide samples with
different resistances have to be made individually. Varying the film thickness
in the deposition process was the most effective way to control the sample
resistance, and it turned out that small changes around 200\AA \ yielded the
range of few k$\Omega$ to 100M$\Omega$ (at T=4K).

Achieving continuous thin-films of this material is a challenging undertaking.
This seems to be a problem that plagues other preparation techniques used to
produce thin films of this material \cite{13}. This is the reason for the wide
range of rates and oxygen pressure mentioned above that were tested in
attempting to find the optimal conditions. Variations in the deposition-rates
and partial oxygen pressure had much weaker influence on film structure than
the type of substrate used; for nominal thickness in the range 200$\pm
$20~\AA ~the yield of continuous films using (sodium-oxide-rich)
microscope-slides was around 80\% whereas it was less than 5\% (2 in 41
deposition runs) for the SiO$_{\text{2}}$ layer thermally grown on
%TCIMACRO{\TEXTsymbol{<}}%
%BeginExpansion
$<$%
%EndExpansion
100%
%TCIMACRO{\TEXTsymbol{>} }%
%BeginExpansion
$>$
%EndExpansion
Si-wafers. Much better yield of physically continuous specimen were obtained
on both type of substrates by co-depositing Tl$_{\text{2}}$O$_{\text{3}}$ with
$\simeq$6\%~Au evaporated from a Knudsen source. The gold deposition rate was
adjusted independently from that of the Tl$_{\text{2}}$O$_{\text{3}}$. The
Tl$_{\text{2}}$O$_{\text{3-x}}$:Au films turned out to be less sensitive to
the type of substrate, and they were also more stable against agglomeration
and grain-growth than pure films. Grains as large as 2$\mu$m could be seen by
optical microscopy in Tl$_{\text{2}}$O$_{\text{3-x}}$ samples that were
heat-treated at T$\geq$450K. This agglomeration also affected their optical
transmission in the visible as shown in Fig.~1 comparing Au-doped with undoped
specimen. Susceptibility to agglomeration and the associated grain-growth is
the main impediment in getting electrically continuous thin films of the clean
material.
%TCIMACRO{\FRAME{ftbpFU}{3.4255in}{2.6104in}{0pt}{\Qcb{(color online) Optical
%transmission of 200\AA ~thick Tl$_{\text{2}}$O$_{\text{3-x}}$ and
%Tl$_{\text{2}}$O$_{\text{3-x}}$:Au deposited on glass-slide. Full labels - as
%deposited at room temperatures. Empty labels - after 30 seconds heat-treatment
%at T=490K. Note the large change in transmission for the Tl$_{\text{2}}%
%$O$_{\text{3-x}}$ film as compared with the milder effect in the
%Tl$_{\text{2}}$O$_{\text{3-x}}$:Au sample.}}{}{fig_1.eps}%
%{\special{ language "Scientific Word";  type "GRAPHIC";
%maintain-aspect-ratio TRUE;  display "PICT";  valid_file "F";
%width 3.4255in;  height 2.6104in;  depth 0pt;  original-width 10.1311in;
%original-height 7.6981in;  cropleft "0";  croptop "1";  cropright "1";
%cropbottom "0";  filename 'Fig_1.eps';file-properties "XNPEU";}}}%
%BeginExpansion
\begin{figure}[ptb]%
\centering
\ifcase\msipdfoutput
\includegraphics[
height=2.6104in,
width=3.4255in
]%
{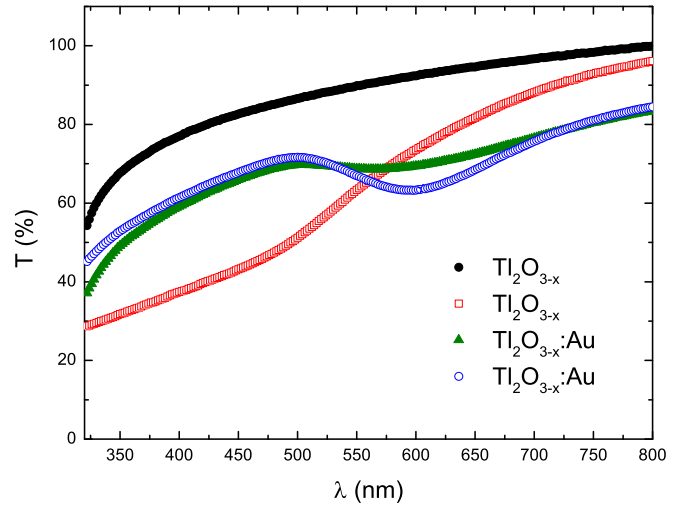}%
\else
\includegraphics[
height=2.6104in,
width=3.4255in
]%
{C:/TEX/TlO/graphics/Fig_1__1.pdf}%
\fi
\caption{(color online) Optical transmission of 200\AA ~thick Tl$_{\text{2}}%
$O$_{\text{3-x}}$ and Tl$_{\text{2}}$O$_{\text{3-x}}$:Au deposited on
glass-slide. Full labels - as deposited at room temperatures. Empty labels -
after 30 seconds heat-treatment at T=490K. Note the large change in
transmission for the Tl$_{\text{2}}$O$_{\text{3-x}}$ film as compared with the
milder effect in the Tl$_{\text{2}}$O$_{\text{3-x}}$:Au sample.}%
\end{figure}
%EndExpansion

Transmission-electron-microscopy (TEM) was used to characterize the films
composition and microstructure (employing the Tecnai F20 G2 equipped with an
energy dispersive analyzer). Samples studied by TEM were deposited unto
carbon-coated Cu grids under the same conditions as those used for transport
measurements. TEM micrographs and electron-diffraction patterns of
Tl$_{\text{2}}$O$_{\text{3-x}}$ and Tl$_{\text{2}}$O$_{\text{3-x}}$:Au films
are shown in Fig.~2 and Fig.~3 respectively. The main difference between the
two types of the Tl$_{\text{2}}$O$_{\text{3-x}}$ films is the preponderance of
smaller and mid-size grains in the Tl$_{\text{2}}$O$_{\text{3-x}}$:Au films,
but both show a rather wide distribution of grain-sizes. This should be
compared with the space-filling, mosaic structure of In$_{\text{2}}%
$O$_{\text{3-x}}$ films [see e.g., Fig.~1 in \cite{14}).
%TCIMACRO{\FRAME{ftbpFU}{3.4255in}{3.4255in}{0pt}{\Qcb{Bright-field micrograph
%of an as-deposited Tl$_{\text{2}}$O$_{\text{3-x}}$ film (nominal thickness
%200\AA ). Note that a typical grain-size rarely exceeds $\simeq$150nm. The
%inset shows the associated diffraction pattern. Note that the bare areas in
%the film (white regions in the micrograph) tend to be adjacent to relatively
%large grains, a consequence of grain growth in a deposit with strong
%self-affinity. Film continuity involves amorphous (or micro-crystalline
%phase)}}{}{fig_2.eps}{\special{ language "Scientific Word";  type "GRAPHIC";
%maintain-aspect-ratio TRUE;  display "PICT";  valid_file "F";
%width 3.4255in;  height 3.4255in;  depth 0pt;  original-width 14.5873in;
%original-height 14.5873in;  cropleft "0";  croptop "1";  cropright "1";
%cropbottom "0";  filename 'Fig_2.eps';file-properties "XNPEU";}}}%
%BeginExpansion
\begin{figure}[ptb]%
\centering
\ifcase\msipdfoutput
\includegraphics[
height=3.4255in,
width=3.4255in
]%
{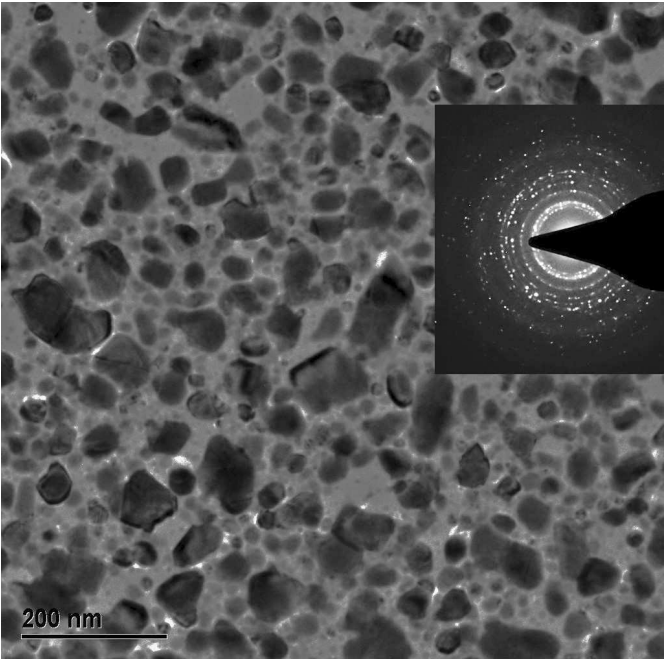}%
\else
\includegraphics[
height=3.4255in,
width=3.4255in
]%
{C:/TEX/TlO/graphics/Fig_2__2.pdf}%
\fi
\caption{Bright-field micrograph of an as-deposited Tl$_{\text{2}}%
$O$_{\text{3-x}}$ film (nominal thickness 200\AA ). Note that a typical
grain-size rarely exceeds $\simeq$150nm. The inset shows the associated
diffraction pattern. Note that the bare areas in the film (white regions in
the micrograph) tend to be adjacent to relatively large grains, a consequence
of grain growth in a deposit with strong self-affinity. Film continuity
involves amorphous (or micro-crystalline phase)}%
\end{figure}
%EndExpansion%
%TCIMACRO{\FRAME{ftbpFU}{3.4255in}{3.5293in}{0pt}{\Qcb{Bright-field micrograph
%of a typical $\simeq$200\AA \ thick Tl$_{\text{2}}$O$_{\text{3-x}}$:Au film.
%The inset shows the associated diffraction pattern. The magnification is the
%same as in Fig.~2.}}{}{fig_3.eps}{\special{ language "Scientific Word";
%type "GRAPHIC";  maintain-aspect-ratio TRUE;  display "PICT";
%valid_file "F";  width 3.4255in;  height 3.5293in;  depth 0pt;
%original-width 14.2431in;  original-height 14.6751in;  cropleft "0";
%croptop "1";  cropright "1";  cropbottom "0";
%filename 'Fig_3.eps';file-properties "XNPEU";}}}%
%BeginExpansion
\begin{figure}[ptb]%
\centering
\ifcase\msipdfoutput
\includegraphics[
height=3.5293in,
width=3.4255in
]%
{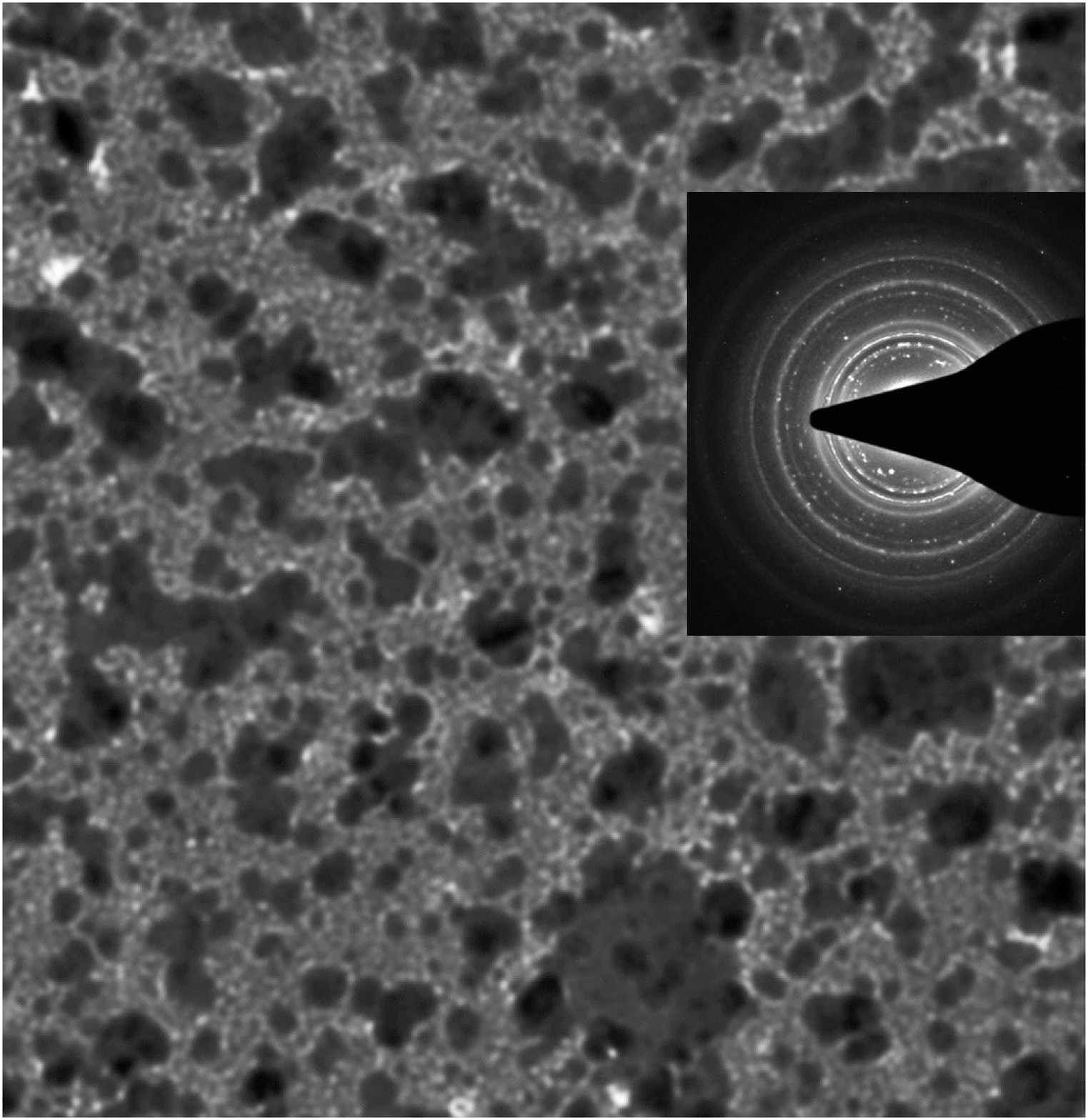}%
\else
\includegraphics[
height=3.5293in,
width=3.4255in
]%
{C:/TEX/TlO/graphics/Fig_3__3.pdf}%
\fi
\caption{Bright-field micrograph of a typical $\simeq$200\AA \ thick
Tl$_{\text{2}}$O$_{\text{3-x}}$:Au film. The inset shows the associated
diffraction pattern. The magnification is the same as in Fig.~2.}%
\end{figure}
%EndExpansion
\-

The diffraction patterns of both types of structures are dominated by the
larger grains giving spotty rings which mask the presence of an amorphous
component in the as-deposited films. The set of rings could be indexed for 8
of the allowed rings according to the bcc bixbyite structure of the compound
with lattice spacing a=10.53$\pm$0.02\AA \ (see table 1).

\begin{center}%
%TCIMACRO{\TeXButton{B}{\begin{table}[tbp] \centering}}%
%BeginExpansion
\begin{table}[tbp] \centering
%EndExpansion%
\begin{tabular}
[c]{|c|c|c|}\hline
\textbf{(hkl)} & \textbf{Observed d (\AA )} & \textbf{Calculated d
(\AA )}\\\hline
(222) & 3.04$\pm$0.01 & 3.040\\\hline
(123) & 2.81$\pm$0.01 & 2.814\\\hline
(400) & 2.64$\pm$0.01 & 2.632\\\hline
(431) & 2.05$\pm$0.01 & 2.065\\\hline
(611) & 1.80$\pm$0.01 & 1.806\\\hline
(622) & 1.59$\pm$0.01 & 1.587\\\hline
(631) & 1.55$\pm$0.01 & 1.553\\\hline
(640) & 1.45$\pm$0.01 & 1.460\\\hline
\end{tabular}%
%TCIMACRO{\TeXButton{caption}{\caption
%{Measured vs. calculated d-spacings of the diffraction rings observed in the TEM micrographs}%
%}}%
%BeginExpansion
\caption
{Measured vs. calculated d-spacings of the diffraction rings observed in the TEM micrographs}%
%EndExpansion
\label{TableKey copy(1)}%
%TCIMACRO{\TeXButton{E}{\end{table}}}%
%BeginExpansion
\end{table}%
%EndExpansion

\end{center}

Energy-dispersive x-ray spectroscopy revealed about 1\% (atomic ) tungsten as
the only contaminant on this level, presumably resulting from spurious
deposition off the e-gun filament, as also noticed in In$_{\text{2}}%
$O$_{\text{3-x}}$ deposition \cite{11}.

In addition to the rings listed in table 1, Tl$_{\text{2}}$O$_{\text{3-x}}$:Au
films showed a fcc pattern of Au grains suggesting that most of the material
segregated out of the Tl$_{\text{2}}$O$_{\text{3-x}}$ matrix. The Au
diffraction pattern was used as a guide to calibrate the camera constant. The
appearance of free Au particles in the Tl$_{\text{2}}$O$_{\text{3-x}}$:Au
films is another aspect in which this material differs from In$_{\text{2}}%
$O$_{\text{3-x}}$ based specimen; in the latter system no sign of free Au is
observed in the diffraction patterns for similar levels of gold addition, and
there are indications that the gold atoms reside in the oxygen vacancies sites.

Samples used for electrical measurements were deposited through rotatable
stainless-steel masks to define their geometry and were typically 1x1mm and
200~\AA \ thick. Prior to sample deposition, gold contacts were evaporated
through another set of masks on the same fixture which had alignment pins that
positioned the contacts on the sample area.

Room temperature Hall-effect measurements were used to determine the
carrier-concentration. This was done with three of the batches prepared for
the transport measurements using samples with a Hall-bar configuration
(six-probe measured by dc). At room temperatures, these films had
sheet-resistances R$_{\square}~$of few k$\Omega~$making them effectively
diffusive, and therefore the Hall-effect may be used as a reliable measurement
of the material carrier-concentration. Essentially the same result for the
carrier-concentration n=(1.9$\pm$0.2)$\cdot$10$^{\text{21}}$cm$^{\text{-3}}$
was found for both Tl$_{\text{2}}$O$_{\text{3-x}}$ and Tl$_{\text{2}}%
$O$_{\text{3-x}}$:Au films. This value for n is larger by 2-3 than the
carrier-concentration found for Tl$_{\text{2}}$O$_{\text{3-x}}$ prepared by
electrodeposition \cite{15} and, interestingly, it is almost two orders of
magnitude larger than that of In$_{\text{2}}$O$_{\text{3-x}}$ \cite{11}. No
difference in transport properties were found between Tl$_{\text{2}}%
$O$_{\text{3-x}}$ and Tl$_{\text{2}}$O$_{\text{3-x}}$:Au films with comparable
sheet-resistance R$_{\square}$. It seems that the only role of Au is to curb
the tendency of the material to agglomerate.

\subsection{Measurement techniques}

Conductivity of the samples was measured using a two-terminal ac technique
employing a 1211-ITHACO current pre-amplifier and a PAR-124A lock-in
amplifier. Most measurements reported below were performed with the samples
immersed in liquid helium at T=4.1K maintained by a 100 liters storage-dewar,
which allowed long term measurements of samples as well as a convenient way to
maintain a stable temperature bath. Unless otherwise indicated, the ac voltage
bias was small enough to ensure linear response conditions (judged by Ohm's
law being obeyed within the experimental error).

The microwaves experiments employed high-power synthesizer (HP8360B) using
power up to 25~dBm ($\approx$316~mW) at the frequency range of 2-6~GHz.
Optical excitation was accomplished by exposing the sample to AlGaAs diode
(operating at $\approx$0.88$\pm$0.05$\mu$m), placed $\approx$15mm from the
sample. The diode was energized by a computer-controlled Keithley 220
current-source. The samples were attached to a probe equipped with calibrated
Ge and Pt thermometers and were wired by triply-shielded cables to BNC
connectors at room temperatures. The effective capacitance of the wires was
typically 10nF allowing the use of 70-75Hz ac technique in most cases.

Fuller details of measurements techniques are given elsewhere \cite{9}.

\section{Results and discussion}

\subsection{Hopping conductivity and current-voltage characteristics}

We start this section by examining the near-equilibrium (conductance) and
steady-state (current-voltage characteristics) properties of the
thallium-oxide films with R$_{\square}$%
%TCIMACRO{\TEXTsymbol{>}}%
%BeginExpansion
$>$%
%EndExpansion
h/e$^{\text{2}}$ (at liquid helium temperature). The only goal in this section
is to demonstrate that the Tl$_{\text{2}}$O$_{\text{3-x}}$samples are in the
strongly-localized regime, and that their near-equilibrium transport
properties cannot be distinguished from those of other hopping systems
(including systems that do not exhibit intrinsic electron-glass effects, e.g.,
lightly-doped semiconductors).

The strongly-localized nature of these films was verified by measuring their
conductance versus temperature dependence G(T) at the temperature range 4-45K.
Figure 4 illustrates this behavior for two of the samples that were used in
this study. Below$\approx$10K, this dependence can be described reasonably
well by G(T)$\propto\exp$[-(T$_{\text{0}}$/T)$^{\text{1/x}}$] with x either 3
or 4. The limited range of temperature and resistance does not allow
distinguishing between these. However, this range is wide enough to rule out
G(T)$\propto\exp$[-(T$_{\text{0}}$/T)$^{\text{1/2}}$] at least for the sample
with the larger resistance in Fig.~4. The form of variable-range-hopping (VRH)
shown in Fig.~4 is often observed in strongly-localized In$_{\text{2}}%
$O$_{\text{3-x}}$ samples \cite{16}. By comparison, over a similar range of
temperatures the hopping law G(T)$\propto\exp$[-(T$_{\text{0}}$%
/T)$^{\text{1/2}}$] is obeyed by thin films of beryllium \cite{10}, and by the
high-carrier concentration version of In$_{\text{x}}$O (with n$\geq
$10$^{\text{21}}$cm$^{\text{-3}}$) \cite{17}.

One should not take these mathematical fits too seriously. It is actually
unclear why these systems should exhibit any specific exponent of VRH; Both,
the ln(G)$\propto$T$^{\text{-1/3}}$ and ln(G)$\propto$T$^{\text{-1/2}}$ VRH
forms are results of models that assume, among other things, single occupation
of localized sites. As remarked elsewhere \cite{16}, this assumption may not
justified in systems with carrier-concentration as high as that of
In$_{\text{2}}$O$_{\text{3-x}}$(let alone in systems with higher n such as Be
and In$_{\text{x}}$O). In these systems there are many charge-carriers within
a localization volume and interactions may modify overlapping electronic states.

There are apparently other factors in hopping conductivity that are not
accounted for by existing theories. A pertinent comment on the uncertainty of
the hopping exponent in realistic systems was made by Brodsky and Gambino on
the basis of their experiments in amorphous Si \cite{18} (where
multi-occupation of localized states is not a concern). Rather, it was the
behavior of the pre-exponential factor that led these authors to remark that
not much weight should be given even to the value of the exponent when fitting
G(T)\ data. The point is that in reality disordered systems are more
complicated than assumed by current theories for hopping conductivity. It is
hardly surprising that important deviations from the predictions of hopping
models are observed in experiments on real samples. Nonetheless, the
exponential temperature dependence of G is a clear indication of the
strongly-localized nature of the used films and that is all we need for the
main conclusions in this paper.
%TCIMACRO{\FRAME{ftbpFU}{3.4255in}{2.4791in}{0pt}{\Qcb{(color online)
%Temperature dependence of the resistance for two of the films used in the
%study (one on each of the type of substrates used). The curves are also
%labeled by their respective activation energy T$_{\text{0}}$. These data may
%be fitted also by R(T)$\propto$exp[(T$_{\text{0}}$/T)$^{\text{1/3}}$] but not
%by R(T)$\propto$exp[(T$_{\text{0}}$/T)$^{\text{1/2}}$].}}{}{fig_4.eps}%
%{\special{ language "Scientific Word";  type "GRAPHIC";
%maintain-aspect-ratio TRUE;  display "USEDEF";  valid_file "F";
%width 3.4255in;  height 2.4791in;  depth 0pt;  original-width 10.3147in;
%original-height 7.4409in;  cropleft "0";  croptop "1";  cropright "1";
%cropbottom "0";  filename 'Fig_4.eps';file-properties "XNPEU";}}}%
%BeginExpansion
\begin{figure}[ptb]%
\centering
\ifcase\msipdfoutput
\includegraphics[
height=2.4791in,
width=3.4255in
]%
{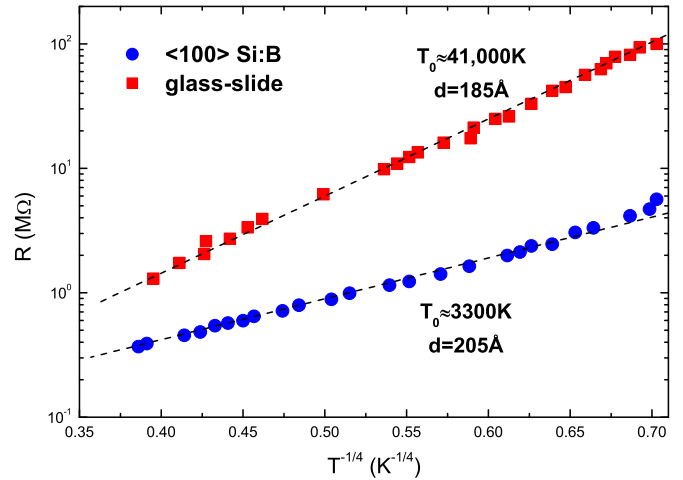}%
\else
\includegraphics[
height=2.4791in,
width=3.4255in
]%
{C:/TEX/TlO/graphics/Fig_4__4.pdf}%
\fi
\caption{(color online) Temperature dependence of the resistance for two of
the films used in the study (one on each of the type of substrates used). The
curves are also labeled by their respective activation energy T$_{\text{0}}$.
These data may be fitted also by R(T)$\propto$exp[(T$_{\text{0}}%
$/T)$^{\text{1/3}}$] but not by R(T)$\propto$exp[(T$_{\text{0}}$%
/T)$^{\text{1/2}}$].}%
\end{figure}
%EndExpansion

A characteristic attribute of hopping systems, which further illustrates the
inadequacy of current VRH models, is their non-Ohmic behavior, which is often
observable at lower fields F than anticipated. The common expectation is that
Ohmic behavior persists up to fields of order k$_{\text{B}}$T/eL where e is
the electron charge and L is the percolation radius \cite{19}. However,
typical values for L determined from the crossover to non-Ohmic behavior often
leads to values of L that are much larger than this length scale
\cite{20,21,22}. The same phenomenon is observed in all our Tl$_{\text{2}}%
$O$_{\text{3-x}}$ and Tl$_{\text{2}}$O$_{\text{3-x}}$:Au samples. Figure 5
shows the dependence of the conductance on the applied voltage for a typical
Tl$_{\text{2}}$O$_{\text{3-x}}$:Au sample.%
%TCIMACRO{\FRAME{ftbpFU}{3.4255in}{2.6201in}{0pt}{\Qcb{(color online) The
%dependence of the conductance on the applied voltage for a typical
%Tl$_{\text{2}}$O$_{\text{3-x}}$:Au film (thickness 200\AA , R$_{\square}%
%$=5.7M$\Omega,$ and lateral dimensions 1x1 mm$^{\text{2}}$) measured at T=4.1K
%using phase-sensitive technique at frequency of 74Hz. The inset shows the
%fractional change of the conductance at the low voltage regime compared with
%the respective result of using a microwave field at f=2.45GHz while measuring
%G using low frequency Ohmic bias (see text).}}{}{fig_5.eps}%
%{\special{ language "Scientific Word";  type "GRAPHIC";
%maintain-aspect-ratio TRUE;  display "PICT";  valid_file "F";
%width 3.4255in;  height 2.6201in;  depth 0pt;  original-width 9.9023in;
%original-height 7.7425in;  cropleft "0";  croptop "1";  cropright "1";
%cropbottom "0";  filename 'Fig_5.eps';file-properties "XNPEU";}}}%
%BeginExpansion
\begin{figure}[ptb]%
\centering
\ifcase\msipdfoutput
\includegraphics[
height=2.6201in,
width=3.4255in
]%
{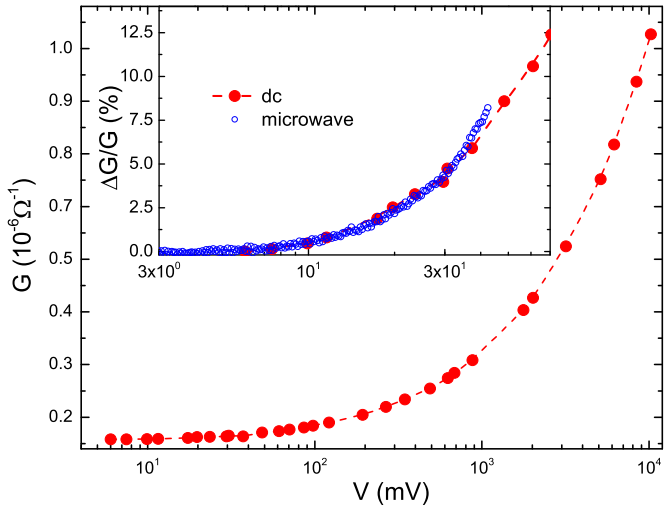}%
\else
\includegraphics[
height=2.6201in,
width=3.4255in
]%
{C:/TEX/TlO/graphics/Fig_5__5.pdf}%
\fi
\caption{(color online) The dependence of the conductance on the applied
voltage for a typical Tl$_{\text{2}}$O$_{\text{3-x}}$:Au film (thickness
200\AA , R$_{\square}$=5.7M$\Omega,$ and lateral dimensions 1x1 mm$^{\text{2}%
}$) measured at T=4.1K using phase-sensitive technique at frequency of 74Hz.
The inset shows the fractional change of the conductance at the low voltage
regime compared with the respective result of using a microwave field at
f=2.45GHz while measuring G using low frequency Ohmic bias (see text).}%
\end{figure}
%EndExpansion

The inset to Fig.~5 is a zoom-in on the low voltage part of the main G(V)
curve taken by a low-frequency ac technique. It is hard to assign a precise
value for the voltage above which Ohmic behavior is no longer obeyed in the
G(V) curve as the deviation is gradual. A fair estimate based on the data
shown in the inset to Fig.~5 would be V$\simeq$30mV where the deviation from
Ohmic behavior is well above the experimental error. Using L$\simeq
$bk$_{\text{B}}$T/eV (b is the sample length) gives L $\simeq$10$\mu$m, which
is much larger than any reasonable estimate for the percolation radius. This
problem seems to be the rule in VRH conductivity; similarly large values for L
were found by Rosenbaum \textit{et al} \cite{20} in doped-Si, and Grannan
\textit{et al} \cite{22} found in Ge samples doped by nuclear
transmutation.The possibility that these discrepancies arise from the present
of long-rnage potential fluctuations was discussed in \cite{20}, and
experimntal evidence for their presence was found in strongly insulating
In$_{\text{x}}$O and In$_{\text{2}}$O$_{\text{3-x}}$ films \cite{23}.

Finally, the change of the conductance due to application of a microwave field
is compared in Fig.~5 (inset) with the G(V) data taken at a low frequency. In
the latter case G is still measured at low frequency but as function of the
amplitude of a microwave field capacitively coupled to the sample \cite{24}.
This microwave enhanced conductivity is another way to modify the hopping
conductance and it is essentially a non-Ohmic effect generic to VRH systems.
This was shown to exist in several systems including in the lightly-doped GaAs
\cite{24}. It was also pointed out that existing models for non-ohmicity in
the VRH regime fail to give a satisfactory fit to the experimental G(V) in
these systems \cite{24}.

To sum up this part we conclude that thallium-oxide samples with R$_{\square}$%
%TCIMACRO{\TEXTsymbol{>}}%
%BeginExpansion
$>$%
%EndExpansion
h/e$^{\text{2}}$ (at T$\approx$4K) exhibit VRH behavior that is similar to
other strongly-localized systems including some lightly-doped semiconductors.
Their out-of-equilibrium transport features described next, identify them as
intrinsic electron-glasses with extended relaxation times as might be expected
from their high carrier-concentration on the basis of previous empirical
findings. These, as will be discussed below, cannot be accounted for by VRH
models that assume single occupation of localized sites.

\subsection{Out of equilibrium transport properties}

The first set of experiments, shown in Fig.~6, describes the basic relaxation
of the excess conductance following quench-cooling the sample from high
temperatures to liquid helium ambience (Fig.~6a). Fig.~6b\&c show the results
of applying the stress-excitation protocol \cite{25} on the same sample after
it was allowed to equilibrate at T=4.1K for $\approx$28 hours. The stress
protocol involves subjecting the sample to a large `source'-`drain' field,
driving the sample into the non-Ohmic regime, which creates excess phonons and
takes the system out of equilibrium. This can be observed during the stress as
a slow increase of the conductance (Fig. 6b), and when Ohmic conditions are
restored a logarithmic decay of the excess conductance is observed (Fig.~6c).%
%TCIMACRO{\FRAME{ftbpFU}{3.4255in}{2.6955in}{0pt}{\Qcb{(color online) Two basic
%manifestations of electron-glass features: Conductance vs. time of a
%Tl$_{\text{2}}$O$_{\text{3-x}}$:Au sample with R$_{\square}$=5.7M$\Omega$ (a):
%following a quench-cool from T$\approx$110K to T=4.1K. (b): Under the stress
%of a field F=102V/cm for 1400 seconds, and (c) After the voltage across the
%sample is set back to an ohmic value. }}{}{fig_6.eps}%
%{\special{ language "Scientific Word";  type "GRAPHIC";
%maintain-aspect-ratio TRUE;  display "PICT";  valid_file "F";
%width 3.4255in;  height 2.6955in;  depth 0pt;  original-width 10.187in;
%original-height 7.9997in;  cropleft "0";  croptop "1";  cropright "1";
%cropbottom "0";  filename 'Fig_6.eps';file-properties "XNPEU";}}}%
%BeginExpansion
\begin{figure}[ptb]%
\centering
\ifcase\msipdfoutput
\includegraphics[
height=2.6955in,
width=3.4255in
]%
{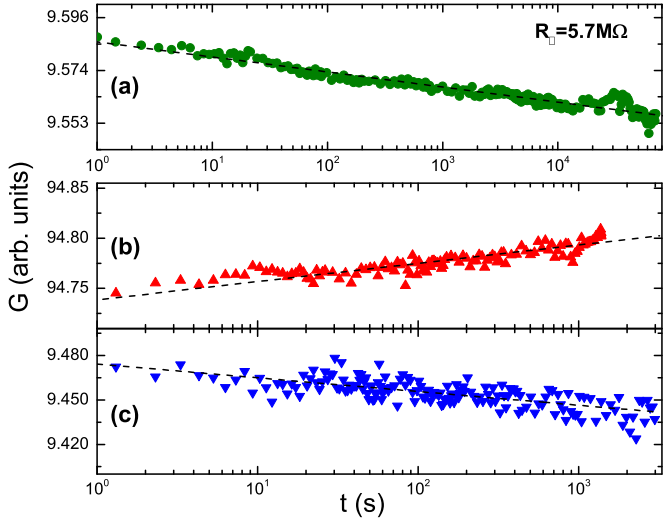}%
\else
\includegraphics[
height=2.6955in,
width=3.4255in
]%
{C:/TEX/TlO/graphics/Fig_6__6.pdf}%
\fi
\caption{(color online) Two basic manifestations of electron-glass features:
Conductance vs. time of a Tl$_{\text{2}}$O$_{\text{3-x}}$:Au sample with
R$_{\square}$=5.7M$\Omega$ (a): following a quench-cool from T$\approx$110K to
T=4.1K. (b): Under the stress of a field F=102V/cm for 1400 seconds, and (c)
After the voltage across the sample is set back to an ohmic value. }%
\end{figure}
%EndExpansion

Another effective way to drive the system far from the equilibrium and observe
the ensuing relaxation is by optical excitation. A typical result using this
protocol is illustrated in Fig.~7 for a Tl$_{\text{2}}$O$_{\text{3-x}}$ film.%
%TCIMACRO{\FRAME{ftbpFU}{3.4255in}{2.5128in}{0pt}{\Qcb{(color online) Typical
%result of an optical excitation protocol using a Tl$_{\text{2}}$%
%O$_{\text{3-x}}$:Au sample with R$_{\square}$=11M$\Omega$ at T=4.1K. Following
%a 3s exposure to infrared light (LED energized by 50mA at 1.5 cm from sample),
%the conductance shows the characteristic logarithmic relaxation law as
%illustrated in the inset.}}{}{fig_7.eps}%
%{\special{ language "Scientific Word";  type "GRAPHIC";
%maintain-aspect-ratio TRUE;  display "PICT";  valid_file "F";
%width 3.4255in;  height 2.5128in;  depth 0pt;  original-width 10.2588in;
%original-height 7.4985in;  cropleft "0";  croptop "1";  cropright "1";
%cropbottom "0";  filename 'Fig_7.eps';file-properties "XNPEU";}}}%
%BeginExpansion
\begin{figure}[ptb]%
\centering
\ifcase\msipdfoutput
\includegraphics[
height=2.5128in,
width=3.4255in
]%
{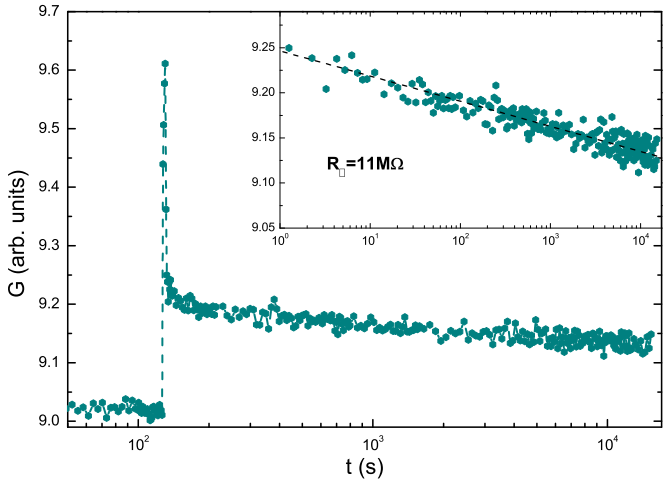}%
\else
\includegraphics[
height=2.5128in,
width=3.4255in
]%
{C:/TEX/TlO/graphics/Fig_7__7.pdf}%
\fi
\caption{(color online) Typical result of an optical excitation protocol using
a Tl$_{\text{2}}$O$_{\text{3-x}}$:Au sample with R$_{\square}$=11M$\Omega$ at
T=4.1K. Following a 3s exposure to infrared light (LED energized by 50mA at
1.5 cm from sample), the conductance shows the characteristic logarithmic
relaxation law as illustrated in the inset.}%
\end{figure}
%EndExpansion

Common to the protocols used in Figs.~6 and 7 is that they can be performed on
films deposited on any insulating substrate and do not necessitate a gate.

This makes it easier to measure the conductance of a high resistance sample. A
nearby metallic gate, a must for the field effect experiments, introduces a
parallel capacitance in the measurement circuit and limits the frequency used
in the ac technique. Also, comparing the results with Tl$_{\text{2}}%
$O$_{\text{3-x}}$ samples configured with a gate helps to verify that the
glassy effects are not due to artifacts associated with trapped charge or
other defects in the insulating layer that separates the sample from the gate
(this appears to be a problem in some semiconductors). However, as was
emphasized elsewhere \cite{6}, the prime signature for an intrinsic
electron-glass is a memory-dip which can only be tested via the field-effect
technique. A major effort was therefore dedicated to produce gated structure
for field effect experiments. Five Tl$_{\text{2}}$O$_{\text{3-x}}$:Au films
with sheet resistances R$_{\square}$ between 800k$\Omega$ to 30M$\Omega$
configured with a gate for field-effect experiments. The MD for the highest
R$_{\square}$ film tested in a MOSFET configuration is shown in Fig.~8.%
%TCIMACRO{\FRAME{ftbpFU}{3.4255in}{2.4108in}{0pt}{\Qcb{(color online) The MD
%for a Tl$_{\text{2}}$O$_{\text{3-x}}$:Au sample with R$_{\square}%
%$=43.9M$\Omega,$ thickness$\approx$195\AA \ and lateral dimensions of
%1.1x1.6mm$^{\text{2}}$ (LxW). The gate voltage was swept in this case from
%V$_{\text{g}}$=-6.3V to V$_{\text{g}}$=+6.4V through the V$_{\text{g}}$=0V
%point where the sample was allowed to equilibrate for 31 hours at T=4.1K. The
%dashed line depicts the slope of the thermodynamic field-effect.}}%
%{}{fig_8.eps}{\special{ language "Scientific Word";  type "GRAPHIC";
%maintain-aspect-ratio TRUE;  display "PICT";  valid_file "F";
%width 3.4255in;  height 2.4108in;  depth 0pt;  original-width 10.5001in;
%original-height 7.3699in;  cropleft "0";  croptop "1";  cropright "1";
%cropbottom "0";  filename 'Fig_8.eps';file-properties "XNPEU";}}}%
%BeginExpansion
\begin{figure}[ptb]%
\centering
\ifcase\msipdfoutput
\includegraphics[
height=2.4108in,
width=3.4255in
]%
{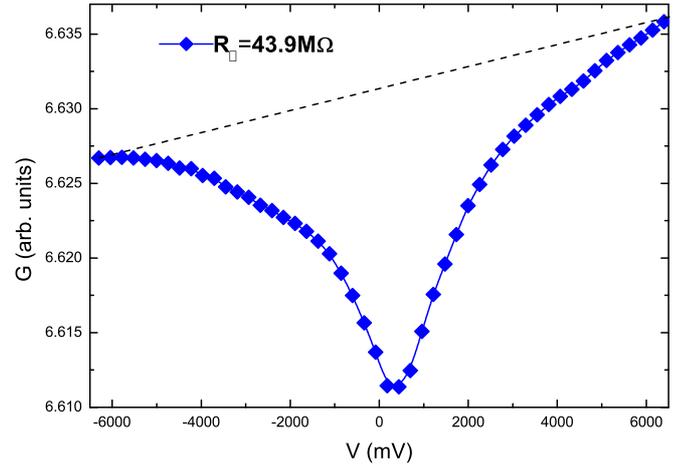}%
\else
\includegraphics[
height=2.4108in,
width=3.4255in
]%
{C:/TEX/TlO/graphics/Fig_8__8.pdf}%
\fi
\caption{(color online) The MD for a Tl$_{\text{2}}$O$_{\text{3-x}}$:Au sample
with R$_{\square}$=43.9M$\Omega,$ thickness$\approx$195\AA \ and lateral
dimensions of 1.1x1.6mm$^{\text{2}}$ (LxW). The gate voltage was swept in this
case from V$_{\text{g}}$=-6.3V to V$_{\text{g}}$=+6.4V through the
V$_{\text{g}}$=0V point where the sample was allowed to equilibrate for 31
hours at T=4.1K. The dashed line depicts the slope of the thermodynamic
field-effect.}%
\end{figure}
%EndExpansion

As in other electron-glasses, the width $\Gamma$ of the MD turns out to be
independent of R$_{\square}$ over the range studied \cite{9}. Note that
$\Gamma$ is comparable to that of the electron-rich version of In$_{\text{x}}%
$O that has a similar value of carrier-concentration n as our Tl$_{\text{2}}%
$O$_{\text{3-x}}$:Au films, as can be seen in Fig.~9a. The significance of
this observation will be discussed later.
%TCIMACRO{\FRAME{ftbpFU}{3.4255in}{2.6104in}{0pt}{\Qcb{(color online)
%Field-effect results for three samples all with the same MOSFET configuration
%(using Si:B as the gate), and after 24 hours of equilibrating at T=4.1K under
%V$_{\text{g}}$=0. The arrows show the measure of the width $\Gamma$ defined at
%the half-height of the MD as was done in \cite{7}. (a) Comparing the MD of the
%Tl$_{\text{2}}$O$_{\text{3-x}}$:Au film with R$_{\square}$=5.7M$\Omega$ (same
%as in Fig.~5\&6) with a In$_{\text{x}}$O film with R$_{\square}$=8.3M$\Omega.$
%(b) G(V$_{\text{g}}$) for a diffusive Tl$_{\text{2}}$O$_{\text{3-x}}$:Au
%sample (thickness 220\AA , R$_{\square}$=4.1k$\Omega$). }}{}{fig_9.eps}%
%{\special{ language "Scientific Word";  type "GRAPHIC";
%maintain-aspect-ratio TRUE;  display "PICT";  valid_file "F";
%width 3.4255in;  height 2.6104in;  depth 0pt;  original-width 10.2996in;
%original-height 7.8276in;  cropleft "0";  croptop "1";  cropright "1";
%cropbottom "0";  filename 'Fig_9.eps';file-properties "XNPEU";}}}%
%BeginExpansion
\begin{figure}[ptb]%
\centering
\ifcase\msipdfoutput
\includegraphics[
height=2.6104in,
width=3.4255in
]%
{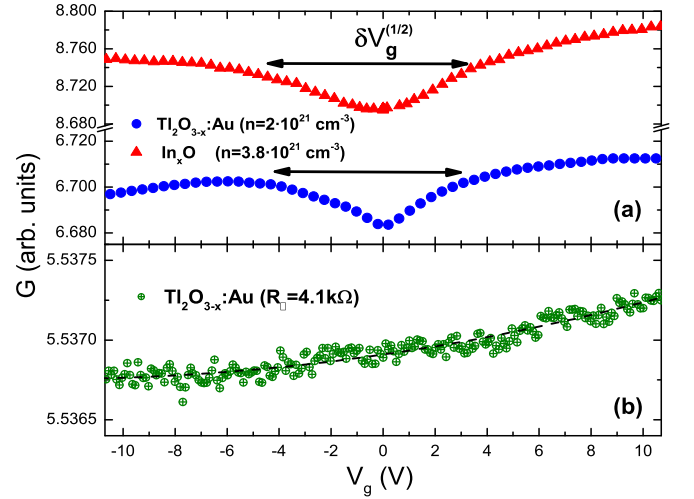}%
\else
\includegraphics[
height=2.6104in,
width=3.4255in
]%
{C:/TEX/TlO/graphics/Fig_9__9.pdf}%
\fi
\caption{(color online) Field-effect results for three samples all with the
same MOSFET configuration (using Si:B as the gate), and after 24 hours of
equilibrating at T=4.1K under V$_{\text{g}}$=0. The arrows show the measure of
the width $\Gamma$ defined at the half-height of the MD as was done in
\cite{7}. (a) Comparing the MD of the Tl$_{\text{2}}$O$_{\text{3-x}}$:Au film
with R$_{\square}$=5.7M$\Omega$ (same as in Fig.~5\&6) with a In$_{\text{x}}$O
film with R$_{\square}$=8.3M$\Omega.$ (b) G(V$_{\text{g}}$) for a diffusive
Tl$_{\text{2}}$O$_{\text{3-x}}$:Au sample (thickness 220\AA , R$_{\square}%
$=4.1k$\Omega$). }%
\end{figure}
%EndExpansion

The presence of a memory-dip in the G(V$_{\text{g}}$) scans is only observable
at T=4K for samples with R$_{\square}\eqslantgtr$200k$\Omega$. Figure 9b
shows, on the same gate-voltage scale, the field effect taken with another
FET-like structure on a Tl$_{\text{2}}$O$_{\text{3-x}}$:Au film with a lower
disorder. This sample had R$_{\square}$%
%TCIMACRO{\TEXTsymbol{<}}%
%BeginExpansion
$<$%
%EndExpansion
h/e$^{\text{2}}$ and, at this temperature is in the diffusive regime which, as
might be expected, shows no sign of a MD. Rather, the G(V$_{\text{g}}$) curve
in Fig.~9b exhibits only the thermodynamic field-effect as a slightly concave
G(V$_{\text{g}}$) curve presumably reflecting the energy dependence of the
density of states $\frac{\partial n}{\partial\mu}$. The curvature of this
curve is typical to a disorder-modified $\frac{\partial n}{\partial\mu
}(\varepsilon)$ at the tail of the conduction band (note that the Fermi energy
associated with n$\simeq$10$^{\text{21}}$cm$^{\text{-3}}$ is near the bottom
of the band). The $\partial$G/$\partial$V$_{\text{g}}$%
%TCIMACRO{\TEXTsymbol{>}}%
%BeginExpansion
$>$%
%EndExpansion
0 is consistent with the material being n-type degenerate semiconductor
\cite{15}.

The thermodynamic contribution to the field-effect is observable whether the
system is glassy or not. For samples in the glassy regime it appears as an
anti-symmetric component in $\partial$G/$\partial$V$_{\text{g}}$. Another way
to see the relative contributions of the non-equilibrium and thermodynamic
components is in the temporal response of the conductance to a one-step change
in V$_{\text{g}}$. An example of such a protocol is illustrated in Fig.~10.
This is another method of driving the system out of equilibrium and observe
the ensuing relaxation.
%TCIMACRO{\FRAME{ftbpFU}{3.4255in}{2.4694in}{0pt}{\Qcb{(color online)
%Excitation by a fast change of the gate voltage (3 seconds swing from
%equilibrium to the target V$_{\text{g}}$). Sample is the same as in
%Figs.~5\&6. Note the difference in the ensuing G(V$_{\text{g}}$) produced by
%the opposite polarity sweeps (taken from the equilibrium V$_{\text{g}}$=0
%point). The difference, marked by the double-headed arrow) is due to the
%contribution of the thermodynamic field-effect. T=4.1K}}{}{fig_10.eps}%
%{\special{ language "Scientific Word";  type "GRAPHIC";
%maintain-aspect-ratio TRUE;  display "PICT";  valid_file "F";
%width 3.4255in;  height 2.4694in;  depth 0pt;  original-width 10.5569in;
%original-height 7.5855in;  cropleft "0";  croptop "1";  cropright "1";
%cropbottom "0";  filename 'Fig_10.eps';file-properties "XNPEU";}}}%
%BeginExpansion
\begin{figure}[ptb]%
\centering
\ifcase\msipdfoutput
\includegraphics[
height=2.4694in,
width=3.4255in
]%
{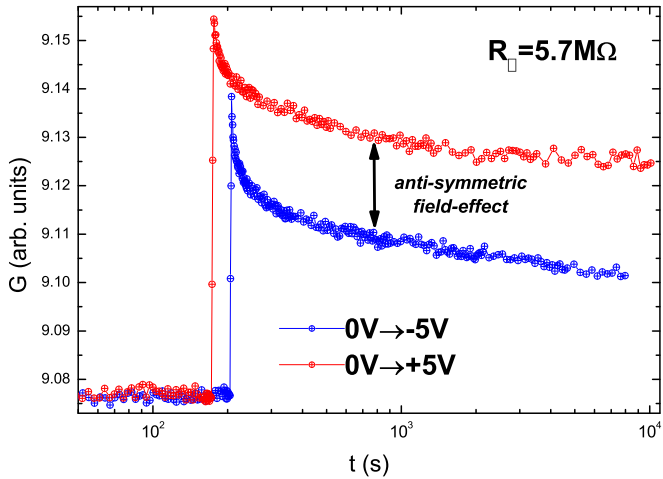}%
\else
\includegraphics[
height=2.4694in,
width=3.4255in
]%
{C:/TEX/TlO/graphics/Fig_10__10.pdf}%
\fi
\caption{(color online) Excitation by a fast change of the gate voltage (3
seconds swing from equilibrium to the target V$_{\text{g}}$). Sample is the
same as in Figs.~5\&6. Note the difference in the ensuing G(V$_{\text{g}}$)
produced by the opposite polarity sweeps (taken from the equilibrium
V$_{\text{g}}$=0 point). The difference, marked by the double-headed arrow) is
due to the contribution of the thermodynamic field-effect. T=4.1K}%
\end{figure}
%EndExpansion

There are two fundamental properties that the memory-dip exhibits. First, its
position on the gate-voltage scale, that in the above plots is centered at
V$_{\text{g}}$=0, depends on the gate-voltage at which the system was allowed
to equilibrate. Secondly, it has a characteristic shape that does not depend
on whether it was recorded by starting the sweep \textit{away} from the
equilibrium V$_{\text{g}}$ towards either polarity or by sweeping
V$_{\text{g}}$ \textit{through} the equilibrium point. These two features,
previously studied in the In$_{\text{2}}$O$_{\text{3-x}}$ and In$_{\text{x}}$O
samples \cite{9}, are illustrated in Fig.~11. The figure shows a set of
recorded events where initially the sample was cooled and equilibrated while
holding the gate-voltage at V$_{\text{g}}$=+2.6V. After equilibration, the two
different sweep protocols were taken to show that the MD is now centered at
+2.6V, and that the shape of the MD is essentially the same for both
protocols. Then, V$_{\text{g}}$ was moved to V$_{\text{g}}$=0V and it was held
there for 24 hours. Subsequent scans of V$_{\text{g}}$ now reveal the MD
centered at the new equilibration position while the old MD has almost
vanished (leaving a small bulge at the right-wing of the `new' MD).
%TCIMACRO{\FRAME{ftbpFU}{3.4255in}{2.4694in}{0pt}{\Qcb{(color online) Four
%gate-voltage sweeps recording G(V$_{\text{g}}$) for a 215\AA \ thick
%Tl$_{\text{2}}$O$_{\text{3-x}}$:Au sample with R$_{\square}$=4.5M$\Omega$ at
%T=4.1K. The first two were taken from V$_{\text{g}}$=+2.6V to both polarities
%(see text) after the sample was equilibrated there for 34 hours and the
%resulting G(V$_{\text{g}}$) traces were spliced to form the plot labeled by
%squares. The two other G(V$_{\text{g}}$) traces were taken as indicated in the
%figure. T-4.1K}}{}{fig_11.eps}{\special{ language "Scientific Word";
%type "GRAPHIC";  maintain-aspect-ratio TRUE;  display "PICT";
%valid_file "F";  width 3.4255in;  height 2.4694in;  depth 0pt;
%original-width 10.5569in;  original-height 7.5855in;  cropleft "0";
%croptop "1";  cropright "1";  cropbottom "0";
%filename 'Fig_11.eps';file-properties "XNPEU";}}}%
%BeginExpansion
\begin{figure}[ptb]%
\centering
\ifcase\msipdfoutput
\includegraphics[
height=2.4694in,
width=3.4255in
]%
{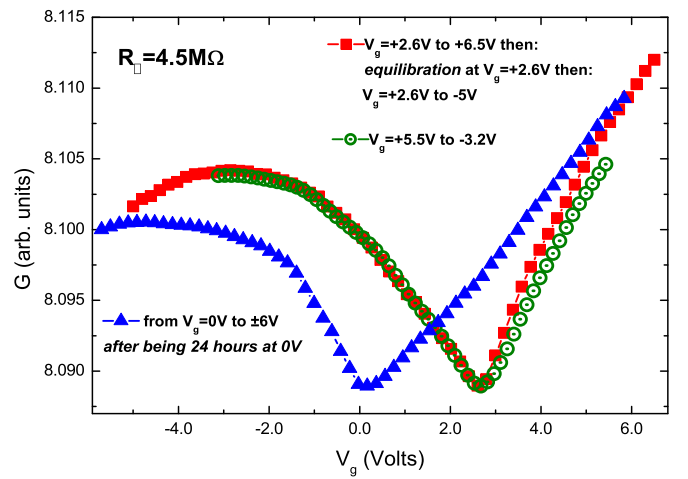}%
\else
\includegraphics[
height=2.4694in,
width=3.4255in
]%
{C:/TEX/TlO/graphics/Fig_11__11.pdf}%
\fi
\caption{(color online) Four gate-voltage sweeps recording G(V$_{\text{g}}$)
for a 215\AA \ thick Tl$_{\text{2}}$O$_{\text{3-x}}$:Au sample with
R$_{\square}$=4.5M$\Omega$ at T=4.1K. The first two were taken from
V$_{\text{g}}$=+2.6V to both polarities (see text) after the sample was
equilibrated there for 34 hours and the resulting G(V$_{\text{g}}$) traces
were spliced to form the plot labeled by squares. The two other G(V$_{\text{g}%
}$) traces were taken as indicated in the figure. T-4.1K}%
\end{figure}
%EndExpansion

The non-equilibrium transport features described above are remarkably similar
to those reported in other intrinsic electron-glasses in all aspects and, in
particular, in the phenomenology associated with the MD. To see how the width
of the MD fits the dependence on carrier-concentration we give in Fig.~12 the
width $\Gamma$ of other non-granular electron-glasses measured to date. In the
figure $\Gamma$ is given in energy units calculated from the experimentally
measured width $\delta$V$_{\text{g}}^{\text{(1/2)}}$ (see e.g., Fig.$~$9a) by:
$\Gamma=\frac{\delta V_{\text{g}}^{\text{(1/2)}}C}{e\frac{\partial n}%
{\partial\mu}(E_{\text{F}})\lambda}$.$~$Here C is the sample-to-gate
capacitance (per unit area), $\frac{\partial n}{\partial\mu}$($\varepsilon
_{\text{F}}$) the material DOS at the Fermi energy, e is the electron-charge,
and $\lambda$ is the screening-length. $\frac{\partial n}{\partial\mu}%
$($\varepsilon_{\text{F}}$) and $\lambda$ were estimated from the material
carrier-concentration n using free-electron formulae. The values of $\Gamma$
obtained in this way are shown in figure 12 as a function of the inter-carrier
distance
%TCIMACRO{\TEXTsymbol{<}}%
%BeginExpansion
$<$%
%EndExpansion
r%
%TCIMACRO{\TEXTsymbol{>}}%
%BeginExpansion
$>$%
%EndExpansion
$\propto$n$^{\text{-1/3}}$. The carrier-concentration n is based on Hall
effect measurement performed for each sample included in the figure. The
single data point for In$_{\text{2}}$O$_{\text{3-x}}$ is actually the averaged
value for 24 measured samples. Similarly, the data for beryllium is the
average of 9 samples \cite{10}, the data for In$_{\text{2}}$O$_{\text{3-x}}%
$:Au is based on 5 samples, and the data for Tl$_{\text{2}}$O$_{\text{3-x}}%
$:Au is based on the 5 samples of the present study. Amorphous indium-oxide,
In$_{\text{x}}$O, is the only EG material studied to date where a study of
samples over a wide range of n has been achievable. However, all the other
systems follow the same systematic dependence on n as the In$_{\text{x}}$O samples.

It is noteworthy that every granular metal that has been tested show
electron-glass effects that are quite similar to those of the Anderson
insulators in Fig.~12. The studied granular materials include granular-Pb
\cite{26}, granular-Al \cite{27}, and discontinuous films of Au \cite{28}, Ni
and Ag \cite{29}. In the hopping regime, all these systems exhibit a
memory-dip with $\Gamma$ values that are at the high end of the group depicted
in Fig.~12. These are not shown in the figure because their
carrier-concentration was not measured. Nonetheless, it is plausible that n in
such systems is on the high side, which lends further support to the
correlation between carrier-concentration and $\Gamma$.

In an earlier work where $\Gamma$(n) was studied (Fig.~4 in \cite{8}), the
possible connection between the MD and the Coulomb-gap was considered. Some of
the reservations raised in \cite{8} with this line of interpretation were
later relaxed \cite{6}. The correlation between $\Gamma,$ that by now has been
fortified by adding more samples and three more Anderson insulators, is
suggestive of a generic mechanism. This lead us to re-examine the connection
between the Coulomb-gap and the memory-dip.%
%TCIMACRO{\FRAME{ftbpFU}{3.4255in}{2.4059in}{0pt}{\Qcb{(color online) The
%typical width (see text for definition) of the memory-dip $\Gamma$ as function
%of the carrier-concentration for five different materials. The dashed line
%depicts the relation: $\Gamma\propto n^{\text{-1/3}}$ consistent with a 1/r
%energy scale. Note that MD width defined here may be smaller than the energy
%scale characterizing the interactions (see reference 6 for full discussion).
%All data are at T=4.1K}}{}{fig_12.eps}{\special{ language "Scientific Word";
%type "GRAPHIC";  maintain-aspect-ratio TRUE;  display "PICT";
%valid_file "F";  width 3.4255in;  height 2.4059in;  depth 0pt;
%original-width 11.0296in;  original-height 7.7274in;  cropleft "0";
%croptop "1";  cropright "1";  cropbottom "0";
%filename 'Fig_12.eps';file-properties "XNPEU";}}}%
%BeginExpansion
\begin{figure}[ptb]%
\centering
\ifcase\msipdfoutput
\includegraphics[
height=2.4059in,
width=3.4255in
]%
{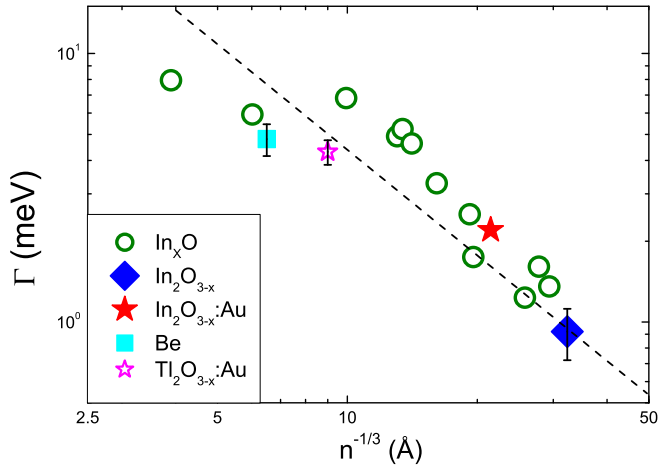}%
\else
\includegraphics[
height=2.4059in,
width=3.4255in
]%
{C:/TEX/TlO/graphics/Fig_12__12.pdf}%
\fi
\caption{(color online) The typical width (see text for definition) of the
memory-dip $\Gamma$ as function of the carrier-concentration for five
different materials. The dashed line depicts the relation: $\Gamma\propto
n^{\text{-1/3}}$ consistent with a 1/r energy scale. Note that MD width
defined here may be smaller than the energy scale characterizing the
interactions (see reference 6 for full discussion). All data are at T=4.1K}%
\end{figure}
%EndExpansion

The Coulomb gap (CG), first discussed by Pollak \cite{30} and Srinivasan
\cite{31} is a consequence of the loss of screening in the disordered medium,
which makes the interaction long-ranged. Efros and Shklovskii has shown how
this leads to a soft gap in the single-particle DOS of the system. There are
two features of the CG that are reminiscent of the memory-dip main
characteristics. First, the CG is a depression in the single particle density
of states N($\varepsilon$) with its minimum pinned to the Fermi energy
$\varepsilon_{\text{F}}$. This would be a natural explanation for the
observation that the MD is centered at the gate-voltage at which the system
equilibrated as well as for the two-dip-experiment (Fig. 11 above, \cite{8,9}
and references therein). Secondly, the correlation between the MD width and
carrier-concentration (Fig.~7) is also in accord with the CG scenario; it is
expected that the energy scale due to interactions will be larger in systems
with higher carrier-concentration.

The possible relevance of the Coulomb-gap to the non-equilibrium features
observed in the intrinsic electron-glasses was considered by Clare Yu
\cite{32,33} and by Lebanon and M\"{u}ller \cite{34,35}. Yu has adopted the
Efros-Shklovskii approach according to which the Coulomb-gap determines the
system conductance. This allowed her to reproduce the evolution with time of
the MD following a quench from high temperatures. Lebanon and M\"{u}ller,
based on similar ideas, developed a model that accounts for many of the
observed electron-glass features. In particular, they gave a plausible
interpretation to the temperature dependence of the MD shape \cite{34}. This
is one of the features that distinguishes the MD of the electron-glass from
the cusp that appears in the structural two-level-system \cite{36,37}; the
shape of the cusp in the structural two-level system does not change with
temperature whereas the MD shape is very sensitive to temperature \cite{6,9}.

The consistency of the theoretical approaches \cite{32,33,34,35} with the main
features of the experiments is encouraging. It does look as a step in the
right direction. At the same time, there are questions of fundamental nature
that should be addressed. The first question is the subtle issue of the
connection between a single-particle DOS and any equilibrium (or near
equilibrium) quantity like conductivity. Patrick Lee has commented on this
issue with regard to the Thomas-Fermi screening length and the use of the
Einstein relation \cite{38}. He pointed out that in both cases it is the
thermodynamic DOS $\frac{\partial n}{\partial\mu}$ that is the relevant quantity.

The single-particle DOS is relevant for processes associated with insertion
(removal) of a particle into (from) the system while no relaxation of other
particles takes place, which would be the case in tunneling or photoemission
events. By contrast, the field effect is usually conceived in the same
category as conductance in this respect.

In the current situation we are in an uncharted territory; the process of
sweeping the gate voltage involves inserting (removing) particles while
"\textit{some} fast things have happened but \textit{some} slow things have
not", which means: things are changing\textit{ even as we sweep}. Under these
nonequilibrium conditions it is not clear whether the use of the Einstein
relation is justified. One may argue that if the sweep duration is short
relative to the relaxation time (yet keeping $\Delta$V$_{g}$/$\Delta$t small
enough to minimize the risk of structural changes that may be caused by the
sweep), the situation is no different than the near-equilibrium conditions
encountered in real-life situations. There are no indication from our
non-equilibrium field-effect experiments to contradict this point of view. In
fact, the dependence of the MD magnitude on the gate-voltage scan-rate seems
to be consistent with it; it is found (Fig.~4 in \cite{6}) that the MD
magnitude increases with the scan-rate with the same logarithmic law that
controls the time-relaxation of the excess conductance. It may be said that
G(V$_{\text{g}}$) is a snap-shot of a DOS made observable by a scan that
occurred while some particles in the system are effectively frozen. Whether
these crude ideas are justified or not must await a proper theory for
field-effect experiments in out-of-equilibrium situations.

Even if it is the single-particle DOS that modulates G(V$_{g}$), it is not
clear what functional form should be taken for it. The problem is not only the
multitude of forms derived for the Coulomb gap by different authors
\cite{39,40,41,42,43,44,35}, it is rather the basic model used; essentially
all Coulomb-gap models assume \textit{single}-occupation of sites while the
one features common to \textit{all} systems that exhibit MD is that their
localized states contain \textit{many} particles even for strong disorder.
There were several attempts to include quantum effects to assess their
modifications to the Coulomb gap \cite{40,41,45,46,47,48,49,50}. There were
also some attempts to consider double-occupation of the localized states that
already introduced new effects that are manifested in magneto-conductance
measurements \cite{51}. However, to our knowledge, no treatment of the Coulomb
gap for the multi-occupation scenario has been attempted.

The observation that a memory-dip is seen only in electronic systems
containing many particles in a localization volume even when \textit{strongly}
localized \cite{52} may be an important clue to the dynamics of the
electron-glass. Multiply-populated localized-sites are essentially small
(disordered) metals. Tunneling transitions between these mesoscopic elements
would induce a many-body shake-up process which may reduce transition
probabilities due to the Anderson orthogonality-catastrophe (AOC) \cite{53}.
While this is probably a small effect for transport, it may be significant
factor in determining the relaxation rate. The space-energy charge
organization, building up the Coulomb-gap from the out of equilibrium random
distribution, involves many more sites than those participating in the
transport. Moreover, many-sites \textit{simultaneous} transitions are required
to approach the lowest energy configuration, while transport is essentially a
single-particle process. The necessity of many-particle transitions for
achieving the ground state was realized by Baranovski \textit{et al} in
computer simulations \cite{54}.

In sum, we presented data on the structure and transport properties of
thallium-oxide films. In the Anderson insulating regime these films exhibit
hopping conductivity and current-voltage characteristics that are commonly
found in other insulating systems at low temperatures. Their non-equilibrium
transport features are in line with those previously reported in other
electron-glasses. In particular, they exhibit a memory-dip that is
characterized by an energy width that fits the empirical relation relating the
width to the carrier-concentration of the material. This feature makes it a
challenge to account for the memory-dip by a mechanism that links the glassy
effects with 1/f noise \cite{55}.

A common feature shared by all the systems that exhibit intrinsic
electron-glass behavior is high carrier-concentration. This presumably is what
distinguishes them from the lightly-doped semiconductors. Near the transition
to the diffusive regime (or when the dielectric constant of the medium is very
large), the electron-electron interaction will be much smaller than the
disorder and therefore the element of frustration, crucial for glassy
behavior, may not be fulfilled. Both strong disorder and comparably strong
interactions are presumably needed for the glassy phase.

Empirically, Anderson insulators with higher carrier-concentration tend to
exhibit long relaxation times. Granular metals present a special case of
strongly-localized systems with high carrier-concentration and it is
remarkable that they systematically exhibit MD and long relaxation times. Note
that the other feature they share with the systems in Fig. 12 is many
electrons per `localization volume'. The empirical evidence in favor of the
conjecture that multi-occupation is the reason for the long relaxation times
is compelling enough to warrant a serious theoretical effort to explore the
possibility that many-body effects are involved.

To further test the empirical connection between carrier concentration and
relaxation time, it is desirable to include more materials in the experimental
study. Both low and high carrier-concentration should be tested. On the
theoretical side, there is the challenge of dealing with the consequences to
hopping transport due to multi-occupation of electronic states, and the
connection between the single-particle DOS and conductivity in non-equilibrium situations.

\begin{acknowledgement}
I am indebted to Dr. Tsirlina for drawing my attention to the special
qualities of thallium-oxide and for sharing her experience in sample
preparation. This research has been supported by a grant administered by the
Israel Academy for Sciences and Humanities.
\end{acknowledgement}


\begin{thebibliography}{99}
\bibitem {52}Near the transition to the diffusive regime (or when the
dielectric constant of the medium is very large), the electron-electron
interaction will be much smaller than the disorder and therefore the element
of frustration, crucial for glassy behavior, may not be fulfilled.
\end{thebibliography}
\end{document}